\documentclass[%
 reprint,
 amsmath,amssymb,
 aps,
floatfix,
longbibliography,
]{revtex4-1}

\usepackage{graphicx}
\usepackage{dcolumn}
\usepackage{bm}

\usepackage{multirow}
\usepackage{array}

\begin{document}

\preprint{APS/123-QED}

\title{Magnetic dipole ordering in resonant dielectric metasurfaces}

\author{Vladimir~R.~Tuz$^{1}$}
\author{Pengchao~Yu$^2$}
\author{Victor~Dmitriev$^{3}$}
\author{Yuri~S.~Kivshar$^{4}$}  
\affiliation{$^1$State Key Laboratory of Integrated Optoelectronics, College of Electronic Science and Engineering, International Center of Future Science, Jilin University, 2699 Qianjin Street, Changchun 130012, China}
\affiliation{$^2$College of Physics, Jilin University, Changchun 130012, China}
\affiliation{$^3$Electrical Engineering Department, Federal University of Para, PO Box 8619, Agencia UFPA, CEP 66075-900 Belem, Para, Brazil}
\affiliation{$^4$Nonlinear Physics Center, Australian National University, Canberra ACT 2601, Australia}  

\date{\today}

\begin{abstract}
Artificial magnetism at optical frequencies can be realized in metamaterials composed of periodic arrays of subwavelength elements, also called {\em ``meta-atoms''}.  Optically-induced magnetic moments can be arranged in both unstaggered structures, naturally associated with {\em ferromagnetic} (FM) order, or staggered structures, linked correspondingly to {\em antiferromagnetic} (AFM) order. Here we demonstrate that such magnetic dipole orders of the lattices of meta-atoms can appear in low-symmetry Mie-resonant metasurfaces where each asymmetric dielectric (non-magnetic) meta-atom supports a localized trapped mode. We reveal that these all-dielectric resonant metasurfaces possess not only strong optical magnetic response but also they demonstrate a significant polarization rotation of the propagating electromagnetic waves at both FM and AFM resonances. We confirm these findings experimentally by measuring directly the spectral characteristics of different modes excited in all-dielectric metasurfaces, and mapping near-field patterns of the electromagnetic fields at the microwave frequencies.

\end{abstract}



\maketitle
\section{\label{intro}Introduction}

Natural materials exhibit negligible magnetism at optical frequencies because the direct effects of optical magnetic fields on matter are much weaker than the effects of electric fields~\cite{Landaushitz_book_8}. A lack of strong optical magnetism in natural materials has motivated many researchers to search for various strategies for achieving a strong magnetic response in specially engineered nanostructures and metamaterials. Even being made of non-magnetic constituents, metamaterials can possess many characteristics of magnetic materials being driven by optically-induced magnetic moments. For example, their effective permeability tensor differs from the unity tensor because of specifically chosen form and arrangement of the subwavelength particles (often called `meta-atoms'). In general, an array of meta-atoms responds to the electromagnetic waves as an effective medium demonstrating a dynamic resonant magnetic response (see a comprehensive review on artificial magnetism in Refs.~\cite{Soukoulis_NatPhotonics_2011, Monticone_JMaterChemC_2014, Kivshar_LowTempPhys_2017}).

Several designs of meta-atoms have been proposed to achieve artificial magnetism. A canonical example of the corresponding meta-atom is a metallic split-ring resonator (SRR). If many of such resonators are arranged in a two-dimensional array, they form a metasurface. In the metasurface illuminated by an external electromagnetic wave, magnetic dipole moments arise from the circular current flow induced in the particles' material. Such a mechanism of attaining the artificial magnetism was first demonstrated at microwave frequencies \cite{Pendry_MTT_1999}, and then it was extended to optics by exploiting the plasmonic resonances of metallic nanoparticles~\cite{Yen_Science_2004,Linden_SCIENCE_2004,Soukoulis_PhysRevLett_2005, Lahiri_OptExpress_2010}. It has been developed further for a variety of non-magnetic plasmonic structures ranging from nanobars and nanoparticle clusters~\cite{Shalaev_OptLett_2005, Fan_Science_2010, fan_NanoLett_2010,Sun_NanoLett_2016} to more complicated systems \cite{Cai_OptExpress_2007, Yuan_OptExpress_2007, Pakizeh_JOSAB_2008}. 

However, it becomes clear that plasmonic metasurfaces based on arrays of metallic SRRs cannot support strong optical magnetism at higher frequencies, especially in the infrared and visible parts of spectrum, due to conduction of metals~\cite{Klein_OptLett_2006}. To achieve the artificial magnetism at higher frequencies, it was proposed to use subwavelength meta-atoms made of high-index low-loss dielectric materials. Unlike two-dimensional materials supporting plasmonic modes \cite{Li_PhysRevB_2014, Li_PhysRevB_2015}, electromagnetic characteristics of all-dielectric resonant metasurfaces originate from the displacement (polarization) currents induced by incident radiation in the dielectric particles. Each particle behaves as a dielectric resonator sustaining a set of Mie-type modes~\cite{Kuznetsov_Science_2016, kruk_acsphotonics_2017, kivshar_2018}, while their ensemble creates a macroscopic resonant response of the entire metasurface. 

The basic physical mechanism of the artificial magnetism in the dielectric particles is the excitation of the corresponding modes of the resonator with a circular flow of the displacement currents. The spectral position of resonances can be tuned by changing the size of particles employed for meta-atoms. This mechanism has been realized in diverse metasurfaces composed of dielectric nanoparticles for their operation in the entire visible spectral range~\cite{Evlyukhin_PhysRevB_2010, Evlyukhin_NanoLett_2012, Kuznetsov_Science_2016}. Unfortunately, such resonances arising from the Mie-type modes in all-dielectric metasurfaces do not always satisfy the demands of practice, since they demonstrate lower degree of the near-field localization than that observed for plasmonic resonances in metallic metasurfaces. 

This issue can be resolved utilizing symmetry protected states in metasurfaces \cite{Koshelev_PhysRevLett_2018}. These states are related to a specific class of eigenmodes which have zero electromagnetic coupling to the field of an irradiating wave. Such modes are referred to as the trapped \cite{Zouhdi_Advances_2003, Fedotov_PhysRevLett_2007} or dark \cite{jain_advoptmater_2015} modes. For instance, in the case of all-dielectric metasurfaces composed of nanodisks, the dipole magnetic mode whose magnetic moment is oriented orthogonally to the metasurface plane behaves as a collective trapped mode, since it cannot be excited by a normally incident plane wave with any linear polarization \cite{Tuz_OptExpress_2018}.

Effectively coupled optically-induced magnetic dipole moments can be arranged in both unstaggered structures, naturally associated with {\em ferromagnetic} (FM) order, or staggered structures, linked correspondingly to {\em antiferromagnetic} (AFM) order. This staggered structure of magnetic dipoles was suggested first for a hybrid metal-dielectric meta-atom composed of a silicon nanoparticle (sphere) and a metallic SRR, operating at the THz frequencies~\cite{Miroshnichenko_ACSNano_2012}. By arranging such hybrid meta-atoms into a two-dimensional lattice, an alternating magnetization in the in-plane directions is obtained due to the magnetic interaction between the dielectric spheres. This concept has been developed further for all-dielectric metasurfaces based on diatomic meta-atoms composed of two types of dissimilar Mie-resonant dielectric particles~\cite{Lepeshov_ACSPhotonics_2018}.

Moreover, when particles are aggregated into clusters, strong nearest-neighbor interactions (coupling) in the metasurface array become dominate, and the mutual magnetization of the neighboring elements can appear in the anti-parallel fashion \cite{Wegener_PhysRevB_2009, Decker_OptLett_2009}. The appearance of such dynamic patterns of induced alternating magnetic moments resembles the AFM order. It turned out that AFM order can be easily achieved in the metallic metasurfaces \cite{born_ApplPhysLett_2014, Singh_AdvMat_2018}, since thin SRRs supporting surface currents can be given a quite complicated shape, while its excitation for optical frequencies with the use of volumetric dielectric resonators supporting displacement currents is not so trivial.

In this paper, we develop a novel approach to optically-induced magnetic ordering in metamaterials and demonstrate that in all-dielectric metasurfaces composed of clusters of identical particles, both resonant FM and AFM orders can be achieved at different frequencies for the same geometry of metasurface. These dissimilar magnetic resonances arise due to the splitting of the trapped modes excited in each meta-atom with broken in-plane symmetry. Remarkably, at the resonant frequencies of both FM and AFM structured orders of magnetic dipoles, the metasurface demonstrates a strong polarization transformation of the transmitted and reflected waves which is of great practical importance. 

\section{Trapped modes in dielectric metasurfaces}

We consider all-dielectric metasurfaces supporting trapped (dark) mode resonances which appear in metasurfaces provided that their meta-atoms possess certain structural asymmetry (recently, such type of modes influenced by the symmetry breaking in resonators were attributed to the phenomenon of bound states in the continuum (BIC) \cite{Koshelev_PhysRevLett_2018, koshelev_2019, Kupriianov_PhysRevApplied_2019}). The mechanism of excitation of trapped modes was first introduced to metasurfaces composed of thin metallic SRRs and their complementary patterns \cite{Fedotov_PhysRevLett_2007, Plum_OptExpress_2009, Khardikov_JOpt_2010}, and then extended on the all-dielectric case \cite{Khardikov_JOpt_2012, jain_advoptmater_2015, Tuz_OptExpress_2018, Cui_acsphotonics_2018, Sayanskiy_PhysRevB_2019}. Remarkably, such metasurfaces provide a strong localization of the electromagnetic fields depending on the degree of asymmetry of the meta-atom. By choosing a proper design of such asymmetric meta-atoms, one can excite a trapped mode with a circular displacement current flow and, thus, achieve an artificial magnetism.

\begin{figure}[t!]
\centering
\includegraphics[width=0.9\linewidth]{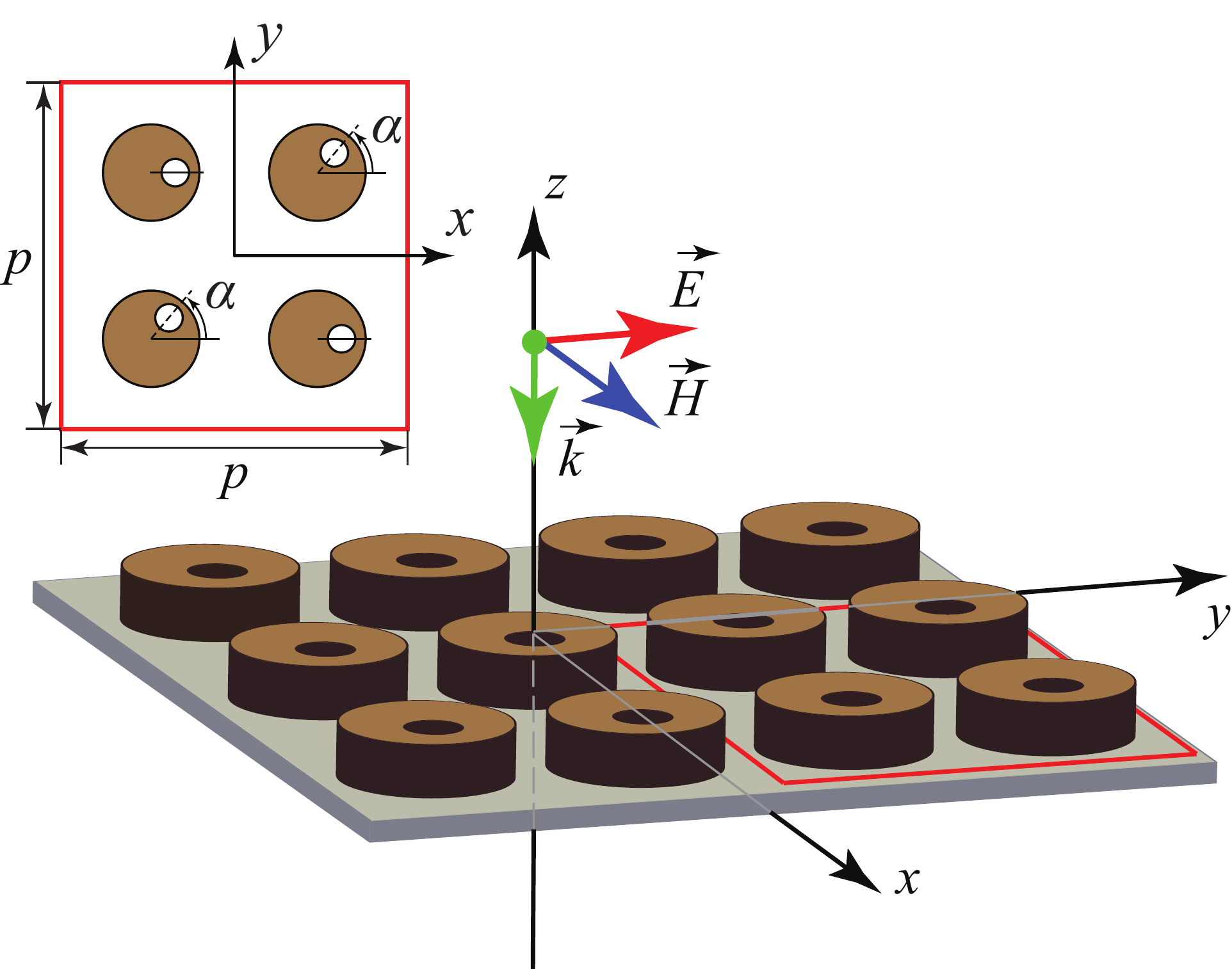}
\caption{Fragment of an all-dielectric metasurface and its complex unit cell (cluster). The structure is under an illumination of a normally incident plane wave.}
\label{fig:sketch}
\end{figure}

In order to construct a dielectric metasurface supporting trapped modes, here we employ a set of identical subwavelength particles (see Fig.~\ref{fig:sketch}). These particles are cylindrical dielectric resonators (disks). The disks radius and thickness are $r_d$ and $h_d$, respectively. They are disposed on a dielectric substrate with the thickness $h_s$. Both disks and substrate are made of non-magnetic materials, whose permittivities are denoted as $\varepsilon_d$ and $\varepsilon_s$, respectively. Each disk is perturbed in-plane by an eccentric through-hole with the radius $r_h$ (the geometry of such disk belongs to the point group $C_s$ having one mirror plane; we use Sch\"oenflies notation for the point groups \cite{barybin2002modern}). These disks are oriented equally and arranged equidistantly into a lattice. In what follows, for definiteness, we assume that the holes are shifted from the disk center on the distance $o_h$ in the direction of the $x$ axis. 

\begin{figure*}[t!]
\centering
\includegraphics[width=1.0\linewidth]{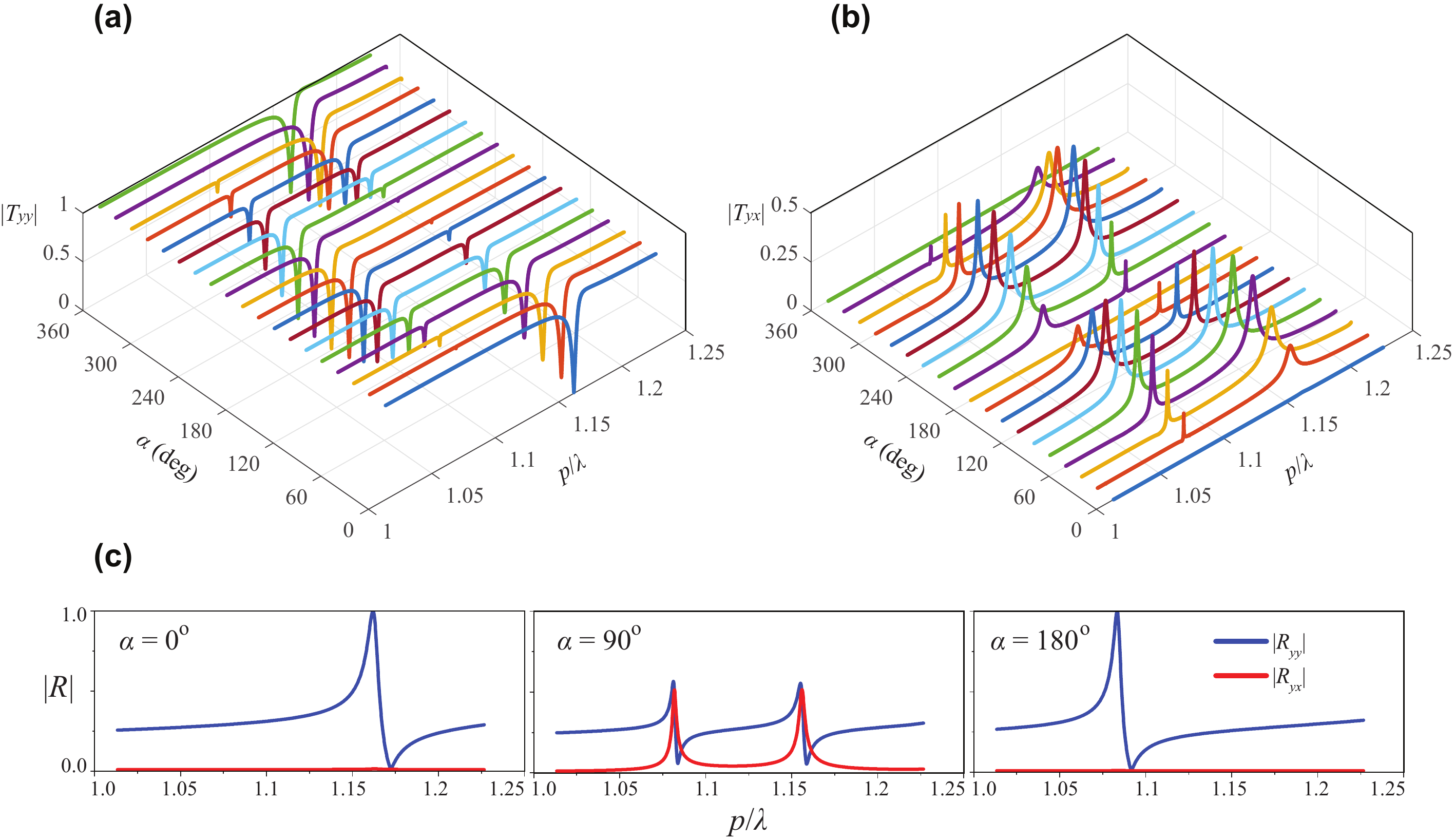}
\caption{(a) Co-polarized, (b) cross-polarized transmission coefficient magnitude, and (c) reflection coefficient magnitude of the ideal (lossless) metasurface as a function of period-to-wavelength ratio and asymmetry parameter $\alpha$; $r_d/p=0.125$, $r_h/r_d =0.375$, $o_h/r_d=0.5$, $h_d/r_d=0.625$, $\varepsilon_d=20.5$, and $\varepsilon_s=1$.}
\label{fig:transmission}
\end{figure*}

Previously it was revealed \cite{Tuz_OptExpress_2018}, that in such a metasurface composed of perturbed resonators, a trapped mode can be excited under the frontal illumination ($\vec k=\{0,0,-k_z\}$), when the vector of electric field of the incident linearly polarized wave does not coincide with the mirror plane of the perturbed resonator. In terms of the eigenwaves of the cylindrical resonator, this trapped mode corresponds to the lowest transverse electric (TE$_{01l}$) mode of the non-perturbed resonator which features the axial (vertical) magnetic moment (the correspondence between eigenwaves and Mie-type modes of a cylindrical resonator see in Tab. 1 of Ref.~\cite{Mongia_1994}). 
Let us further outline in the metasurface a square periodic unit cell (cluster) with the side size $p$ (see insert in Fig.~\ref{fig:sketch}). This cluster encloses four identical disks with holes. Within the unit cell, each disk can be rotated along its axial axis. The final orientation of the holes defines the symmetry of the cluster. Further, we exclude the case when all the holes are oriented randomly, instead, we fix the position of the pair of resonators on one diagonal, while allow rotation of two remaining resonators on a certain angle $\alpha$, where $\alpha \in [0^\circ, 360^\circ)$. With a change in the angle $\alpha$, the symmetry of the square unit cell undergoes transformation $C_{s} \to C_{1} \to C_{s} \to \ldots$ (for the full list of symmetries for the square unit cells composed of four perturbed resonators, see Fig. 3 of Ref.~\onlinecite{yu_JApplPhys_2019}; also the corresponding group subordination scheme (group tree) is given in Fig.~2 of Ref.~\onlinecite{Dmitriev_Metamat_2011}). When the arrangement of disks in the unit cell belongs to the group $C_1$ ($\alpha \ne 0^\circ \wedge \alpha \ne 180^\circ$), the unit cell has no symmetry, and,  thus, the angle $\alpha$ can be considered as an asymmetry parameter of the metasurface. 

Our aim is to investigate the transmission and reflection characteristics of the metasurface illuminated by a normally incident linearly polarized wave. These characteristics can be described by the Jones matrix $\mathbf{J}$, which relates the far-fields of the incident (inc) and outgoing (out) waves as follows:
\begin{equation}
\begin{pmatrix}
E_x^{\textrm{out}} \\
E_y^{\textrm{out}} 
\end{pmatrix} =
\mathbf{J}
\begin{pmatrix}
E_x^{\textrm{in}}\\
E_y^{\textrm{in}} 
\end{pmatrix} =
\begin{pmatrix}
J_{xx} & J_{xy} \\
J_{yx} & J_{yy}
\end{pmatrix}
\begin{pmatrix}
E_x^{\textrm{in}}\\
E_y^{\textrm{in}} 
\end{pmatrix}, 
\label{eq:asymtrans}
\end{equation} 
where the matrix elements $J_{vv'}$ are the co-polarized ($v=v'$) and cross-polarized ($v\ne v'$) components of the transmission ($\mathbf{J}=\mathbf{T}$) or reflection ($\mathbf{J}=\mathbf{R}$) coefficients.

In order to identify the spectral characteristics of the investigated metasurface from the asymmetry parameter, initially we consider that the electric field vector $\vec E^{\textrm{in}}$ of the incident wave is oriented horizontally ($y$-polarization; $\vec E^{\textrm{in}} =\{0,E_y^{\textrm{in}}\}$). 

We perform the numerical simulations of the electromagnetic response of the given metasurface with the use of commercial COMSOL Multiphysics\textsuperscript{\textregistered} finite-element electromagnetic solver. We have modified Model \#15711 from COMSOL Application Gallery \cite{comsol} and use options of the periodic ports and Floquet periodic boundary conditions of the rf module to simulate the infinite two-dimensional array of dielectric resonators. In the first approximation, the ideal (lossless) resonators are considered and the presence of substrate is ignored ($\varepsilon_s=1$). The results of our simulation of the transmission and reflection coefficients are summarized in Fig.~\ref{fig:transmission}.

\begin{figure*}[t!]
\centering
\includegraphics[width=1.0\linewidth]{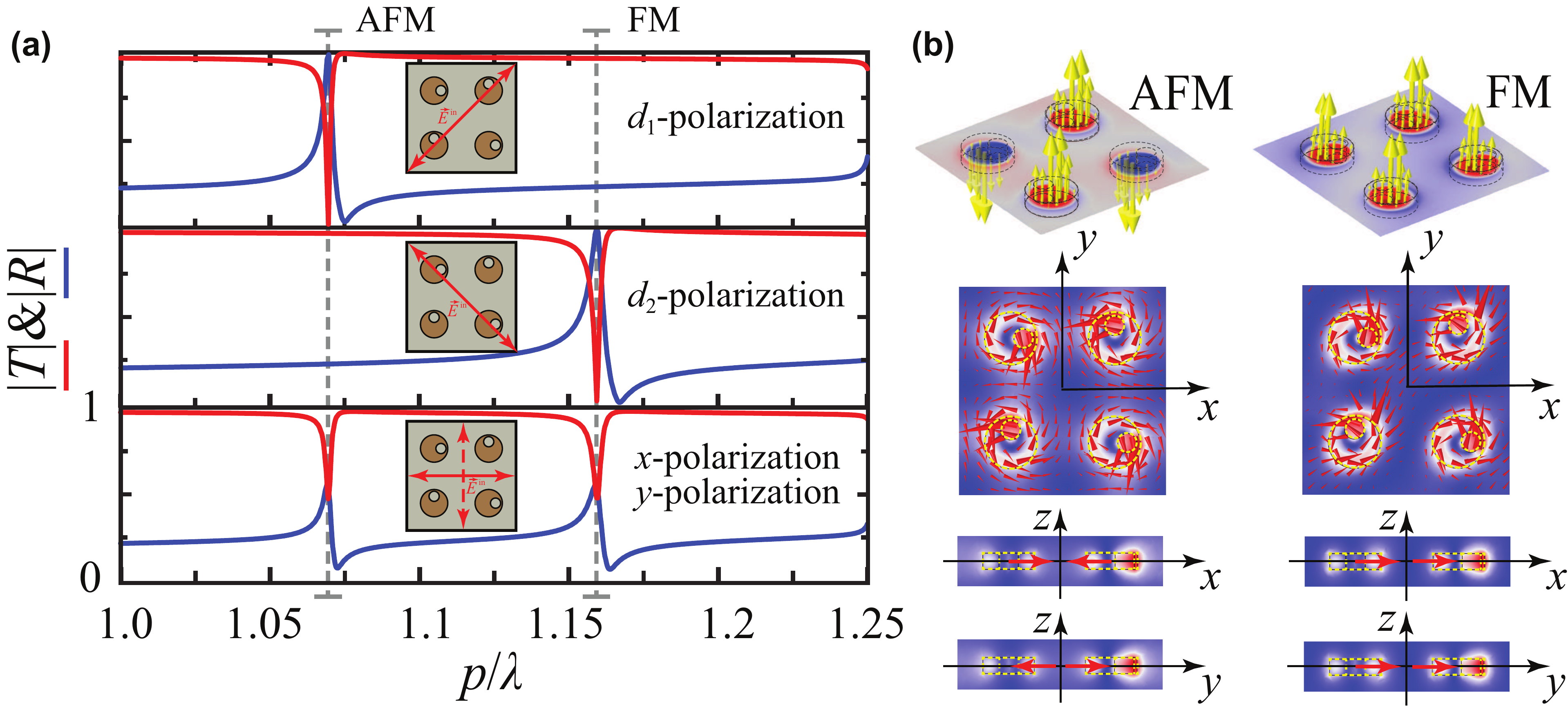}
\caption{(a) Simulated co-polarized transmission and reflection spectra of the ideal (lossless) metasurface excited by a normally incident linearly polarized wave. (b) Distributions of the magnetic field (yellow arrows) and electric field (red arrows) for the symmetric (AFM) and antisymmetric (FM) resonances excited by the $x$-polarized wave. All geometrical and material parameters of the metasurface are the same as in Fig.~\ref{fig:transmission}.}
\label{fig:simulation}
\end{figure*}

In Fig.~\ref{fig:transmission}(a) one can notice, that in the spectral range of interest a particular resonant state appears for each asymmetry parameters $\alpha=0^\circ$ and $\alpha=180^\circ$ when the symmetry of the unit cell corresponds to the group $C_s$. At these values of $\alpha$ there is a difference in the mutual orientations of the perturbed disks in the cluster. In fact, for $\alpha=0^\circ$ and $\alpha=180^\circ$ the plane of symmetry is $y=0$, and $x=0$, respectively. It results in the corresponding resonance appearance at a distinctive frequency. We distinguish them as the high-frequency and low-frequency resonances for $\alpha=0^\circ$ and $\alpha=180^\circ$, respectively. 

Obtained resonances acquire a sharp peak-and-trough (Fano) profile with variation between zero-transmission and zero-reflection which is typical for resonances originated from the trapped modes \cite{Fedotov_PhysRevLett_2007, tuz_PhysRevB_2010, Tuz_EurPhys_2011, Tuz_JOpt_2012}. While the dip in curves corresponds to the maximum of reflection, the peak corresponds to the maximum of transmission. These extremes approach to 0 and 1, respectively, since the material losses are excluded in this simulation. In the rest of the variation range of the asymmetry parameter (i.e., when the symmetry of the unit cell belongs to the group $C_1$), the splitting of the trapped mode into two resonances appears. Remarkably, at both resonant frequencies, the curves of transmission spectra do not vary in the full range between 0 and 1 which may mean that the metasurface performs a polarization transformation. Indeed, the presence of this polarization transformation is confirmed in Figs.~\ref{fig:transmission}(b) and \ref{fig:transmission}(c) for the transmitted and reflected fields, respectively. In what follows our goal is to reveal the characteristics of this polarization transformation from the viewpoint of the artificial magnetism.

\section{\label{theory} Ordering of magnetic dipoles}

In this section, we consider a low-symmetry metasurface that provides the most significant polarization transformation. Its unit cell is constructed of two pairs of perturbed disks with the $\alpha=90^\circ$ difference in the hole rotation between the pairs. As an exciting radiation, we again consider a normally incident linearly polarized plane electromagnetic wave. However, in contrast to the previous consideration, here four different polarization states of the incident wave are under study. They are horizontal ($x$-polarization; $\vec E^{\textrm{in}} =\{E_x^{\textrm{in}},0\}$), vertical ($y$-polarization; $\vec E^{\textrm{in}} =\{0,E_y^{\textrm{in}}\}$), and diagonal polarizations ($d_1$-polarization; $\vec E^{\textrm{in}} =\{E_x^{\textrm{in}},E_y^{\textrm{in}}\}$ and $d_2$-polarization; $\vec E^{\textrm{in}} =\{E_x^{\textrm{in}},-E_y^{\textrm{in}}\}$).

Figure~\ref{fig:simulation}(a) shows simulated transmission and reflection spectra of the metasurface illuminated by a normally incident plane wave for its four different linear polarizations. One can see that in the spectral range of interest, the resonant characteristics for waves of diagonal polarizations are different, while they are the same for the $x$- and $y$-polarized waves. There are the low-frequency resonance and high-frequency resonance in the spectra of the $d_1$- and $d_2$-polarized waves, respectively, and both resonances are in the spectra of the $x$- and $y$-polarized waves. 

To discover the nature of these resonances, we performed the simulation of the electromagnetic near-field distributions at the corresponding resonant frequencies. These simulations are depicted in Fig.~\ref{fig:simulation}(b). From this figure, one can conclude that at the chosen resonant frequencies all resonators forming the cluster are active ones. In each resonator, a particular near-field pattern appears where the arrows of the electric field demonstrate a circular flow in the $x$-$y$ plane. Such a flow produces magnetic moments in each resonator oriented along its axis (along the $z$ axis). These resonances arise from the trapped mode of the perturbed cylindrical resonator \cite{Tuz_OptExpress_2018} and they are not excited entirely in the metasurfaces composed of non-perturbed disks. The trapped mode splits into antisymmetric high-frequency resonance and symmetric low-frequency resonance because there is a coupling between the perturbed resonators in the cluster. While for the antisymmetric resonance the magnetic moments are oriented equally, those for the symmetric resonance are anti-parallel in the adjacent resonators. Therefore, such arrangements of magnetic moments resemble the AFM and FM orders \cite{GARLEA2015203} for the low-frequency and high-frequency resonances, respectively.  

In Fig.~\ref{fig:simulation}(a) one can notice that the resonant curves of transmission and reflection spectra of diagonally polarized waves vary in the range from 0 to 1.  In the chosen frequency band the metasurface under study is almost transparent, whereas at the resonant frequencies of the AFM and FM orders it becomes reflective. Contrariwise, the curves of transmission and reflection spectra of the horizontally and vertically polarized waves vary in the range $0.5-1$ and $0-0.5$, respectively, which is evidence of the sensitivity of the given metasurface to the polarization of the incident wave.

\begin{figure}[t!]
\centering
\includegraphics[width=1.0\linewidth]{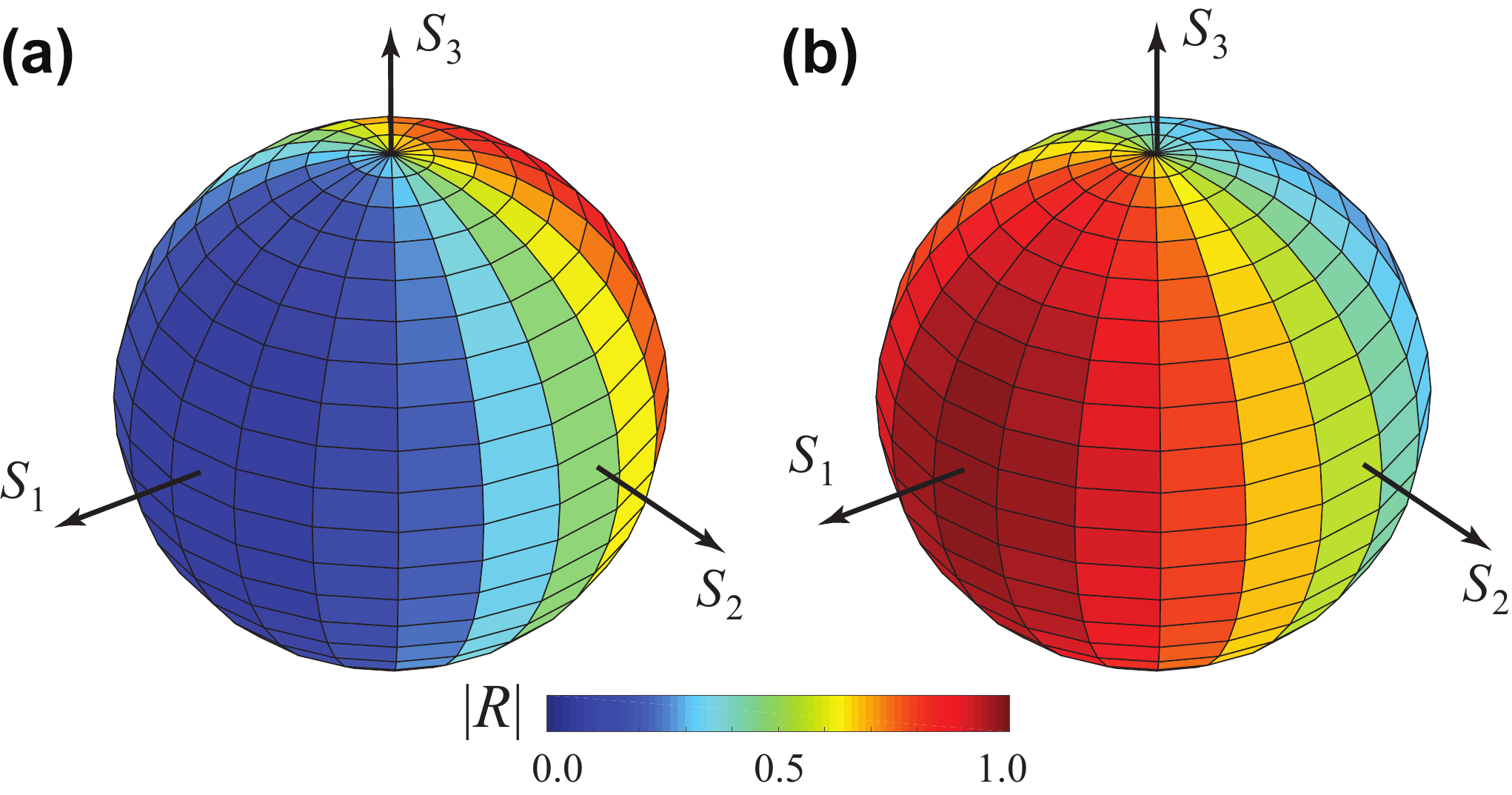}
\caption{Co-polarized reflection coefficient magnitudes for (a) the symmetric (AFM) and (b) antisymmetric (FM) resonances, which are mapped on the surface of a Poincar\'e sphere related to the polarization states of the incident wave. Asymmetry parameter $\alpha$ of the metasurface corresponds to $90^\circ$. All geometrical and material parameters of the metasurface are the same as in Fig.~\ref{fig:transmission}.}
\label{fig:poincare}
\end{figure}

To additionally reveal the polarization sensitivity of the low-symmetry metasurface to any polarization of incident wave we plot the reflection coefficient magnitude of the metasurface as a function of the full Poincar\'e sphere (Fig. \ref{fig:poincare}) at two resonant frequencies. The electric field vector $\vec E^{\textrm{in}}$ of the incident wave is defined by components $E_x^{\textrm{in}}=[0.5(Q+1)]^{1/2}$ and $E_y^{\textrm{in}}=[0.5(Q-1)]^{1/2}\exp(i\beta)$, where $Q\in [-1,1]$ and $\beta \in [-\pi, \pi]$. The axes of the Poincar\'e sphere are labeled as $S_j = U_j/U_0$ ($j=1,2,3$), where $U_0$, $U_1$, $U_2$, and $U_3$ are the Stokes parameters calculated from the components of the reflected electric field in the right-handed orthogonal frame. Linearly polarized states are arranged along the equator of the sphere,while right and left circularly polarized waves are located at its north and south poles, respectively. All other points of the sphere depict elliptically polarized states of the incident wave \cite{Jerrard_JOSA_1954, Collett_ApplOpt_2008}.

\section{\label{exper} Observation of magnetic orders}

We now proceed to the visualization of two above discussed orders of magnetic moments by direct experiment. For plasmonic SRR-based metasurfaces such experiment has been already performed \cite{Diessel_OptLett_2010} by employing scanning near-field optical microscopy. Further we perform both far-field and near-field microwave measurements to demonstrate the manifestation of the AFM and FM orders in our low-symmetry all-dielectric metasurface.

\begin{figure}[t!]
\centering
\includegraphics[width=1.0\linewidth]{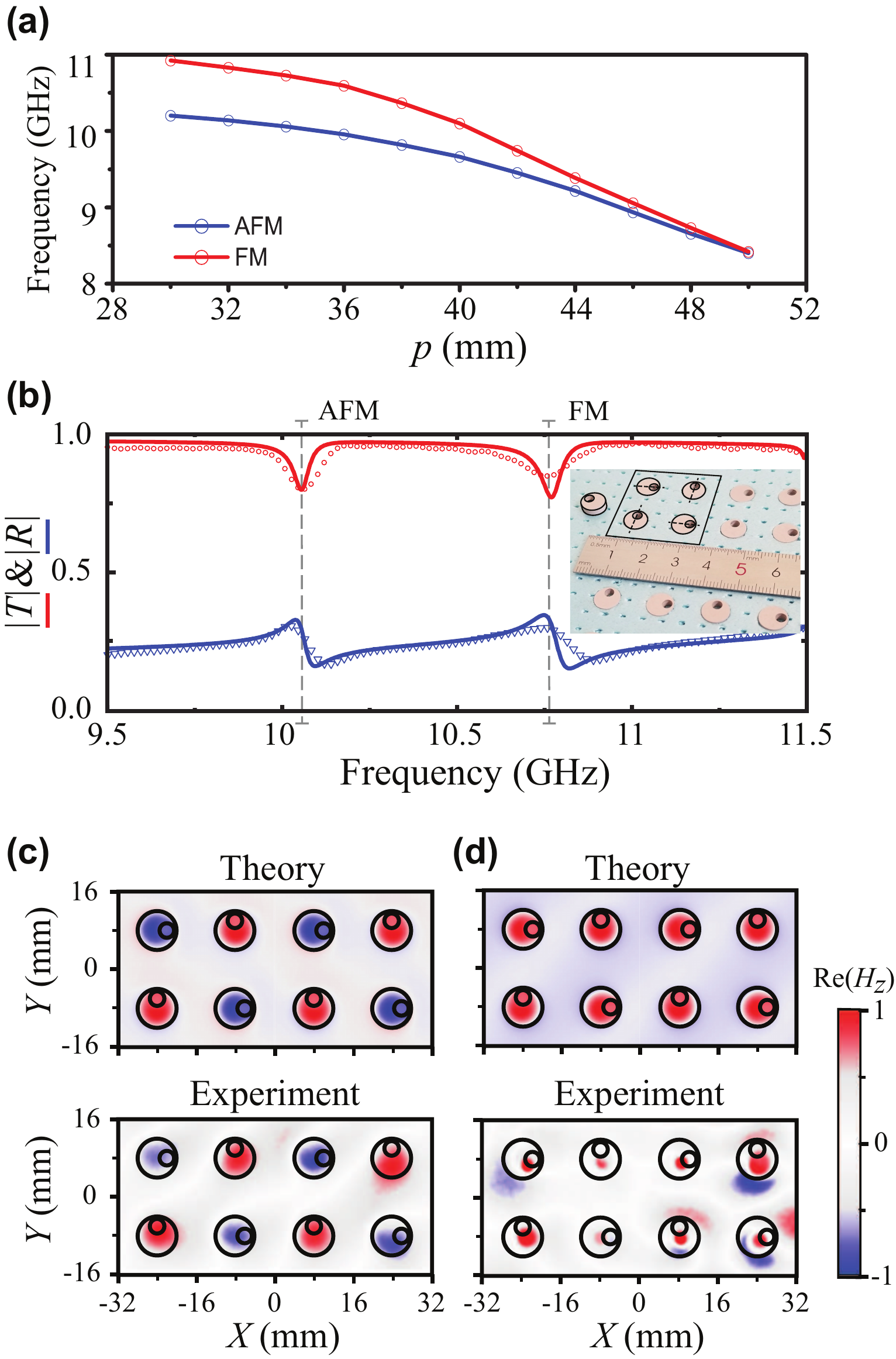}
\caption{(a) Resonant frequencies of the AFM and FM orders versus the lattice period, (b) simulated (solid lines) and measured (scatters) transmission and reflection spectra, and real part of the normal component ($H_z$) of the magnetic near-field (out-of-plane magnetic moments) at the frequencies of (c) AFM and (d) FM resonances excited by the horizontal ($x$-polarized) incident wave. The geometrical and material parameters are: $p = 32$ mm, $\varepsilon_d=20.5$, $\varepsilon_s=1.05$, $\tan\delta_d = 0.015$, $r_d = 4$~mm, $h_d = 2.5$~mm, $h_s = 20.0$~mm, $r_h = 1.5$~mm, and $o_h = 2$~mm.}
\label{fig:experiment}
\end{figure}

Relying on the capabilities of our measurement platform, we choose the microwave frequency range ($9-12$~GHz) for our study and fix the period of the structure to be $p = 32$ mm. In Fig.~\ref{fig:experiment}(a) one can see that at this value of the period, the resonances of interest arise well separated in the chosen frequency range. To assemble a metasurface prototype for experiments, a set of dielectric particles was fabricated. As a dielectric material the Taizhou Wangling TP-series microwave ceramic has been used. The dielectric particles with the sizes mentioned in the caption of Fig.~\ref{fig:experiment} were fabricated with the use of precise mechanical cutting techniques. To arrange them into a metasurface prototype, an array of holes was milled in a custom holder made of rigid foam (Styrofoam) material. The metasurface prototype is composed of $11 \times 11$ clusters (so we used 484 particles in total, and the overall size of the structure is $352 \times 352$ mm$^2$). A photograph of a fragment of the experimental prototype is shown in the insert in Fig.~\ref{fig:experiment}(b).

At the first step, we measured the transmission and reflection spectra of the metasurface prototype. We used a pair of dielectric-lens antennas (HengDa Microwave HD--100LHA250) with a gain of 25~dB over the frequency bandwidth of $8.2-12.4$~GHz. The antennas are specifically oriented to generate and receive a horizontally linearly polarized wave. The prototype is fixed in the middle between the antennas, where a distance between the antennas is 3~m. The antennas are connected to the corresponding ports of the Keysight E5071C Vector Network Analyzer by standard $50$~Ohm coaxial cables. The radiating antenna generates a quasi-plane-wave with required polarization. The beam width at the frequency of 10 GHz is 400 mm with the level of $-3$ dB. With this ratio of the beam width to the overall size of the structure, the structure finiteness has little effect on the spectral characteristics measured. All measurements are performed in an anechoic chamber (details on the experimental setup and measurement technique see in Refs.~\onlinecite{Sayanskiy_PhysRevB_2019, tuz_AdvOptMat_2019}).

The measured transmission and reflection spectra of the metasurface are depicted in Fig.~\ref{fig:experiment}(b). One can see that the curves appeared somewhat smoothed due to the presence of real material losses. We have specified our computational model by accounting material losses existing in ceramic particles. After that, the simulated and measured data are found to be in a good agreement, while their remaining minor deviation can be explained by fabrication imperfection in the actual metasurface prototype. Both resonances predicted by the theory are recognized in the measured spectra, and their resonant frequencies coincide in the measured data and simulations.
 
Next we performed near-field mapping measurements at the corresponding resonant frequencies to obtain a complete confidence in the appearance of specific AFM and FM orders of the magnetic moments. In the measurement setup, a receiving antenna was substituted by an electrically small magnetic dipole probe (a metallic loop with diameter 3 mm) oriented in parallel to the metasurface plane. Two unit cells (eight particles) were included in the scan area. During the measurements the probe was automatically moved over the scan area on a distance of $1$~mm above the prototype surface. The LINBOU near-field imaging system is adapted to perform the movement of probe with a $1$~mm step width in each direction. The real part of the normal component ($H_z$) of the magnetic near-field has been extracted from the measured data and presented in Figs.~\ref{fig:experiment}(c) and \ref{fig:experiment}(d) for the AFM and FM orders, respectively. In this figure, the measured data are supplemented by the corresponding simulated patterns.

The resulting measured and simulated patterns match each other. Obviously, the overall qualitative agreement is very good. At the resonant frequency AFM we observe an alternating pattern of the symmetric mode, whereas at the resonant frequency FM there are regular spots for the antisymmetric mode. It evidences clearly the difference between the AFM and FM orders of magnetic moments originated in the all-dielectric metasurface from the trapped mode splitting. We should note that the measured characteristic of the antisymmetric mode appears somewhat distorted. This is because at this resonant frequency the electromagnetic field is strongly confined inside the dielectric particles (which is specific for the trapped mode excitation). This brings some difficulties for the magnetic near-field probing.

\section{\label{polar} Enhanced polarization effects}

\begin{figure}[t!]
\centering
\includegraphics[width=1.0\linewidth]{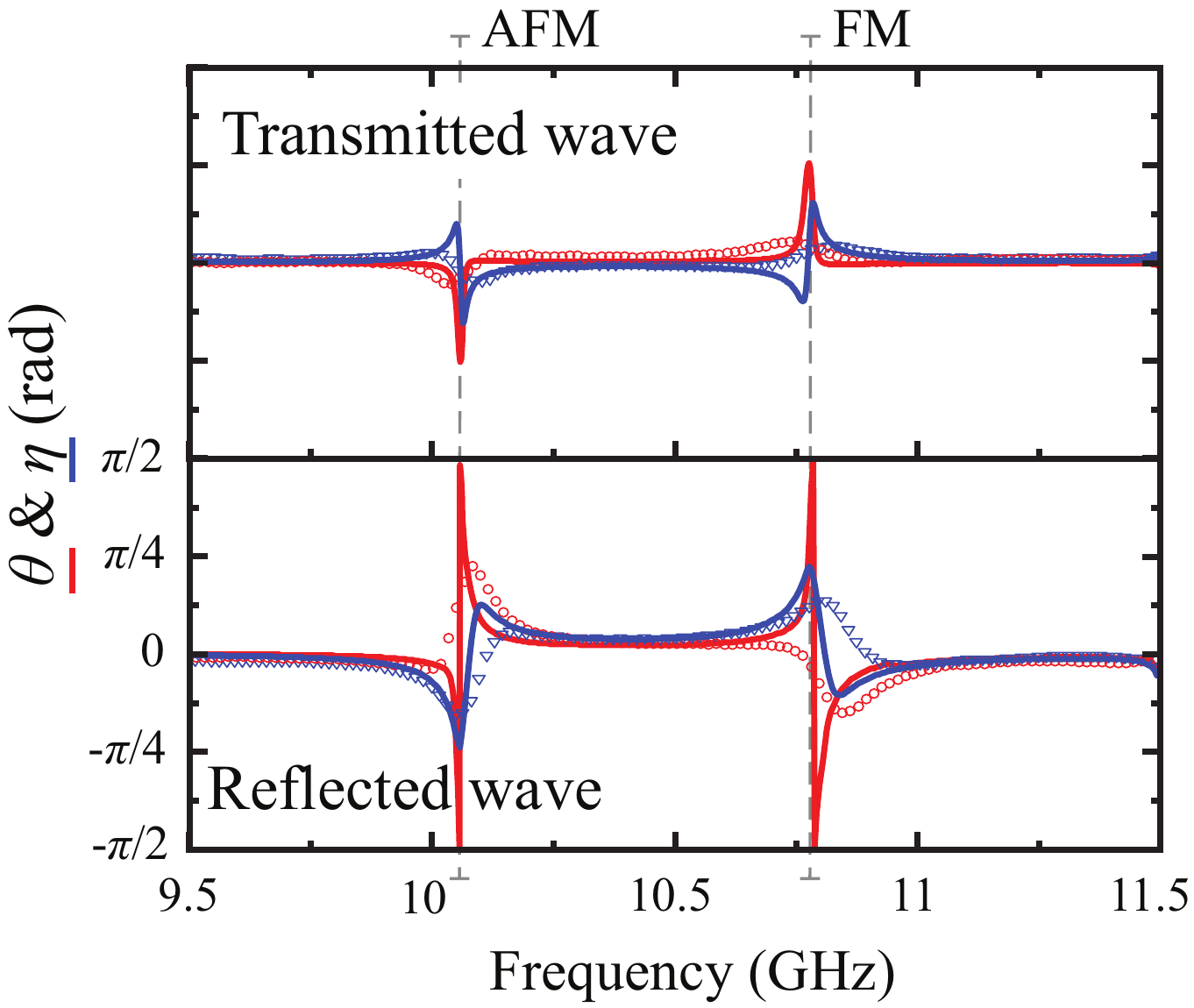}
\caption{Simulated (solid lines) and measured (scatters)  polarization azimuth rotation angle $\theta$ and ellipticity angle $\eta$ of the transmitted and reflected waves. In the simulation an ideal (lossless) metasurface is considered. All geometrical parameters of the actual metasurface are the same as in Fig.~\ref{fig:experiment}.}
\label{fig:ellipticity}
\end{figure}

With the co-polarized and cross-polarized components, the polarization state of both reflected and transmitted fields can be obtained using standard definitions \cite{Collett_1993}: $\tan 2\theta = U_2/U_1$ and $\sin 2\eta = U_3/U_0$, where $\theta$ is the polarization azimuthal angle and $\eta$ is the ellipticity parameter. According to the definition of the Stokes parameters, we introduce the ellipticity $\eta$ so that the field is linearly polarized when $\eta = 0$, and $\eta = -\pi/4$ for left-circular polarization and $+\pi/4$ for right-circular polarization. In   cases with $0 < |\eta| < \pi/4$, the field is elliptically polarized. The frequency dependencies of both the polarization azimuth and ellipticity angle  are summarized in Fig.~\ref{fig:ellipticity}.

Analyzing the properties of the transmitted and reflected fields, we note that generally the metasurface transforms a linearly polarized wave to an elliptically polarized wave in the narrow frequency band around the central resonant frequencies of the AFM and FM orders. There is a stronger polarization conversion in the reflected field where resonant values of the ellipticity angle reach $\pm \pi/4$ (so the reflected field becomes to be circularly polarized). The maximal azimuth rotation angle $\theta$ for the transmitted field is $\pi/4$, whereas for the reflected field it is $\pi/2$. Thus, the metasurface can completely transform the reflected field into an orthogonally polarized wave concerning the incident wave. 

To verify this feature of the metasurface, a set of special measurements of polarization characteristics of the transmitted and reflected fields was conducted. For these experiments, the dielectric-lens antennas were rotated appropriately to measure the co-polarized and cross-polarized components of the transmitted and reflected spectra in the far-field. The results on the polarization azimuth rotation angle and ellipticity angle are derived from the measured data and added in Fig.~\ref{fig:ellipticity} as scatters. One can readily see reasonable agreement between simulations and measurements. Thus the experimental data support our theoretical finding concerning the effect of polarization transformation in the low-symmetry all-dielectric metasurface.

\section{\label{concl}Conclusions}

By employing theoretical methods, numerical simulations, as well as far- and near-field microwave experimental studies, we have revealed and visualized experimentally novel physics of optically-induced magnetism in dielectric metasurfaces. More specifically, we have predicted the ordering of magnetic dipole moments with AFM and FM orders in low-symmetry all-dielectric metasurfaces and observed its experimental manifestation in strong polarization effects. 

Our studies allow making the following conclusions: (i) both symmetric (AFM) and antisymmetric (FM) orders of optically-induced magnetic dipole moments can appear in the low-symmetry all-dielectric metasurface composed of clusters of identical particles, and they do not require magnetic moments of different origin; (ii) both types of the magnetic orders originate from the splitting of the trapped mode whose excitation is supported by the perturbed (asymmetric) particles; (iii) horizontal and vertical linear polarizations acquire a substantial polarization rotation observed experimentally, where the metasurface can perform a complete resonant polarization conversion of the reflected field into an orthogonally polarized wave.   

In the proposed structures, we can gain an additional benefit associated with the excitation of the trapped mode. In particular, under the resonant conditions of the trapped-mode excitation, the electromagnetic field becomes strongly confined inside the structure, and this confinement depends on the degree of asymmetry of individual particles. Therefore, our metasurface can be employed as an efficient flat-optic platform for realizing enhanced light-matter interaction with metasurfaces. 

\section*{\label{ack}Acknowledgments}

V.R.T. and P.Y. acknowledge National Key R\&D Program of China (Grant 2018YFE0119900); V.D. thanks the Brazilian Agency National Council of Technological and Scientific Development (CNPq) for financial support; Y.S.K. acknowledges a support from the Australian Research Council and useful discussions with Boris Luk'yanchuk. 

\bigskip

\bibliography{splitting_trapped}

\end{document}